\documentclass[conference]{IEEEtran}
\IEEEoverridecommandlockouts

\usepackage{cite}
\usepackage{amsmath,amssymb,amsfonts}
\usepackage{multirow}
\usepackage{algorithmic}
\usepackage{graphicx}
\usepackage{textcomp}
\usepackage{tablefootnote}
\usepackage{url}
\usepackage{booktabs}

\usepackage{xcolor}
\def\BibTeX{{\rm B\kern-.05em{\sc i\kern-.025em b}\kern-.08em
    T\kern-.1667em\lower.7ex\hbox{E}\kern-.125emX}}
\begin{document}

\newcommand{\red}[1]{{\color{red}#1}}
\newcommand{\mc}[1]{\mathcal{#1}}
\newcommand{\mr}[1]{\mathrm{#1}}
\newcommand{\mb}[1]{\mathbf{#1}}

\title{State-of-the-art Embeddings with Video-free Segmentation of the Source VoxCeleb Data \\
\thanks{This work was partly supported by project PID2021-125943OB-I00 funded by MCIN/AEI/10.13039/501100011033/FEDER,UE and project PID2024-160789OB-I00 funded by MICIU/AEI/10.13039/501100011033/FEDER,UE and by Tencent AI Lab Rhino-Bird Gift Fund. Collaboration between BUT and Omilia was supported by Horizon 2020 Marie Sklodowska-Curie grant ESPERANTO, No. 101007666. Computing was supported by the Czech Ministry of Education, Youth and Sports through the e-INFRA CZ (ID: 90254)}}
\author{\IEEEauthorblockN{Sara Barahona\IEEEauthorrefmark{1}, Ladislav Mošner\IEEEauthorrefmark{2}, Themos Stafylakis\IEEEauthorrefmark{3}, Oldřich Plchot\IEEEauthorrefmark{2}, Junyi Peng\IEEEauthorrefmark{2}, \\Lukáš Burget\IEEEauthorrefmark{2}, and Jan Černocký\IEEEauthorrefmark{2}}
\IEEEauthorblockA{\IEEEauthorrefmark{1}AUDIAS Research Group, Universidad Autónoma de Madrid, Madrid, Spain}
\IEEEauthorblockA{\IEEEauthorrefmark{2}Brno University of Technology, Faculty of Information Technology, Speech@FIT, Czechia}
\IEEEauthorblockA{\IEEEauthorrefmark{3}Athens University of Economics and Business $|$ Omilia $|$ Archimedes/Athena R.C, Greece}} 


\maketitle

\begin{abstract}
In this paper, we refine and validate our method for training speaker embedding extractors using weak annotations.
More specifically, we use only the audio stream of the source VoxCeleb videos and the names of the celebrities without knowing the time intervals in which they appear in the recording.
We experiment with hyperparameters and embedding extractors based on ResNet and WavLM. We show that the method achieves state-of-the-art results in speaker verification, comparable with training the extractors in a standard supervised way on the VoxCeleb dataset. We also extend it by considering segments belonging to unknown speakers appearing alongside the celebrities, which are typically discarded. Removing the need for speaker timestamps and multimodal alignment, our method unlocks the use of large-scale weakly labeled speech data, enabling direct training of state-of-the-art embedding extractors and offering a visual-free alternative to VoxCeleb-style dataset creation.

\end{abstract}


\section{Introduction}
Extracting speaker embeddings from audio has multiple uses in various speech tasks such as speaker verification, diarization, separation, speech enhancement, voice conversion, multi-speaker automatic speech recognition (ASR), etc. Speaker embedding is a low-dimensional vector that characterizes a person's voice attributes and is extracted from an audio sample typically via large neural networks such as ResNet or TDNN \cite{snyder2018Xvectors,ecapa-tdnn,chung2018voxceleb2}. More recently, with the emergence of self-supervised learning (SSL), the research community has also adopted transformer-based models such as WavLM, HuBERT, or Wav2Vec 2.0\cite{WavLm,hubert,Wav2Vec2} that are either simply used as feature extractors for the mentioned CNN models or combined with a lighter speaker classification backend\cite{mhfa} which allows fast convergence. All of these architectures are typically trained in a supervised way, with training data comprising several thousand speakers, each represented by multiple recordings that contain only their speech.

Creating human-labeled datasets beyond the purpose of the evaluation is prohibitively costly, and the research field has practically moved towards harvesting the data found online while leveraging multiple speech and image-processing technologies to segment it into training samples. One such dataset is VoxCeleb \cite{nagrani17_interspeech, chung18b_interspeech}, which comprises over 2,000 hours of speech recordings from approximately 7,000 celebrities. The speaker interviews are sourced by querying YouTube, selecting audio segments that overlap with the celebrity speaking, using SyncNet for audio-visual synchronization and a ResNet50 for face recognition. By design, this pipeline depends heavily on visual information and excludes any audio segments where the speaker’s face is unclear or outside the frame, resulting in the loss of a substantial amount of data.

SSL has recently emerged as a promising approach for speaker representation learning, aiming to reduce dependence on human-labeled data. Recent state-of-the-art approaches \cite{miara24_interspeech, bingTASL} are based on DINO \cite{Caron2021EmergingPI}, a self-distillation framework where a student network is trained to match the output of a teacher network across multiple augmented iterations of the same utterance. Pseudo-labels are then generated by applying a clustering algorithm over the learned representations. While SSL techniques have significantly closed the gap with supervised methods, they still have not matched its performance and continue to rely on unlabeled, speaker-segmented, face-verified VoxCeleb data, thereby discarding a substantial portion of available audio.


To address this limitation, our previous research \cite{stafylakis22_interspeech} explored training a speaker embedding extractor employing source audio from the VoxCeleb dataset, relying only on recording-level labels while considering the time boundaries where celebrities appear as unknown. Our proposed two-stage method begins by training a speaker embedding extractor on segments derived from clusters obtained via a basic diarization algorithm\cite{broux2018s4d}. 
The trained embedding extractor is then employed to identify which diarized segments belong to the target speaker, which are subsequently used for fully supervised training in the second stage.

\begin{figure*}[h!]
    \centering
    \includegraphics[width=0.8\linewidth]{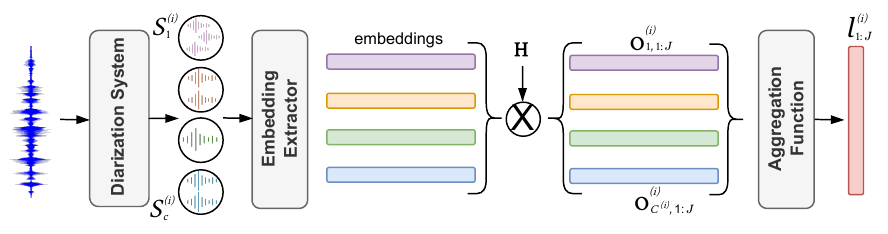}
    \caption{Overview of first-stage training employing \textit{bags} of segments obtained from a basic diarization system. Each mini-batch contains at least one segment from every identified speaker cluster, ensuring speaker representation during training.} 
    \label{fig:stage1}
\end{figure*}
In comparison to our previous work, where we concentrated on formally introducing the method and providing initial results, we now present a comprehensive ablation study of our method on speaker verification task with a newly re-implemented pipeline that is based on PyTorch and the WeSpeaker toolkit\cite{wang2023wespeaker,wang2024advancing}. Motivated by current research with SSL models and the fact that they align well with the philosophy of loosening the labeling requirements of our method, we leverage WavLM features employing Multi-Head Factorized Attention (MHFA) \cite{mhfa} as back-end. Naturally, we also experiment with embedding extractors based on ResNet, as they still constitute the state-of-the-art approach for speaker embedding extraction. We also propose a novel extension to our method that takes into account a portion of segments from unknown speakers, which increases the utilization of the training data. This modification brings further marginal improvements, but more importantly, it shows an interesting and relatively unexplored research direction. Lastly, we assess the impact of segment quality on our method by employing speaker segments generated by a state-of-the-art diarization model from the Pyannote toolkit \cite{Bredin23, Plaquet23}.



\section{Proposed method}
\subsection{Speaker clustering via diarization}
The non-segmented VoxCeleb source data contains information about target celebrities appearing in videos, but lacks crucial temporal annotations regarding their appearance duration and the presence of other speakers. 
Therefore, our initial step involves generating clusters of potential target speakers through diarization. To minimize reliance on annotations, we employ a very basic diarization approach, consisting of (i) a speaker change detection, (ii) Bayesian Information Criterion (BIC)-based hierarchical agglomerative clustering (HAC) with Gaussians on MFCC features for modeling segments, and (iii) Viterbi-based boundary refinement with maximum-likelihood trained Gaussian Mixture Models (GMMs) \cite{Tranter_diar}.

We perform diarization using the SIDEKIT toolkit \cite{broux18_interspeech} with its default hyperparameters, which tend to overestimate the number of speakers in a recording. In our case, this overestimation is desirable, as it helps prevent different speakers from being merged into the same cluster. Therefore, it increases the likelihood that at least one segment in each recording corresponds to the target celebrity during training.

However, this basic approach often results in lower-quality segmentations, which may include long segments with overlapping speakers, background music, or noise. To assess the impact of segmentation quality on our weakly-supervised learning pipeline, we also explore Pyannote.audio \cite{Bredin23, Plaquet23}, a state-of-the-art pre-trained speaker diarization system. This model first employs an end-to-end neural speaker segmentation model \cite{bredin21_interspeech} in short sliding windows, and then local speaker embeddings are extracted for each window. Finally, these embeddings are clustered to assign each local speaker to a global cluster. Although Pyannote has been trained on VoxCeleb data, we do not use it to exploit label information or to match the original training distribution. Instead, our goal is to evaluate whether improved diarization, particularly more accurate segment boundaries, has a significant impact on the performance of our weakly supervised learning pipeline.

 \subsection{Definition of mini-batches}
After the diarization stage, 
each identified audio chunk of the recording $i$ belongs exclusively to one set (i.e., cluster) $\mathcal{S}_c^{(i)}$, where $c$ represents an index of a cluster of audio chunks. Each cluster comprises only chunks belonging to one speaker identity (up to diarization errors). In total, there are $C^{(i)}$ such clusters for the $i$th recording.
As presented in Figure \ref{fig:stage1}, to ensure that each cluster is represented in the mini-batch, we include at least one chunk from each cluster. Therefore, the representation of the $i$th recording in a mini-batch is given by $\{g(S_{c}^{(i)})\}_{c=1}^{C^{(i)}}$ where $S_{c}^{(i)} \in \mathcal{S}_{c}^{(i)}$ and $g(\cdot)$ extracts a fixed-length audio segment. Since the number of clusters differs between recordings, the resulting mini-batches do not have a fixed size and may contain a variable number of segments.

\subsection{First stage: weakly-supervised training}
Following a multi-instance learning strategy \cite{DIETTERICH199731}, our embedding extractor is trained employing a \textit{bag} of segments $\mathcal{S}_c^{(i)}$ from each recording, as defined in the above section. Each segment is processed independently to extract individual embeddings during the feed-forward step. The similarity $o_{c,j}^{(i)}$ between the length-normalized $c$th embedding and the $j$th speaker prototype is computed using the dot product. The $j$th speaker prototype $\mathbf{h}_{j}$ is retrieved from the linear layer of the speaker classification head, defined as $\mathbf{H} = [\mathbf{h}_{1}, \mathbf{h}_{2}, \ldots, \mathbf{h}_{J}]$. 
Considering all clusters, we can collectively represent segment-level similarities as $\phi_j^{(i)} = \{ o_{c,j}^{(i)}\}_{c=1}^{C^{(i)}}$.

To obtain recording-level similarities $\{ l^{(i)}_j \}_{j =1}^{J}$, we aggregate the segment-level similarities from all segments within the same recording. While average pooling could be used for this purpose, it will cause all segments from the same recording to converge towards the prototype, potentially reducing discriminative power. Instead, we use max pooling as aggregating function, which considers only the segment with the highest similarity score and ignores the others. However, because gradients are propagated only through the most similar segment, it might be hard to train effectively during the warm-up phase. To address this, we also considered its soft version, the log-sum-exp (LSE) function, which is smoothed by a temperature parameter $\tau>0$ as follows:

\begin{equation}
l_j^{(i)} = f_{LSE}(\phi_{j}^{(i)};\tau) 
= \tau \log \frac{1}{C^{(i)}} \sum_{c=1}^{C^{(i)}} \exp \left( \frac{1}{\tau} o_{c,j}^{(i)}\right)
\end{equation}
These recording-level similarities are referred to as logits, since they are the values passed to the SoftMax function. Their values range between $[-1, 1]$, making the Additive Angular Margin (AAM) \cite{aam_loss} loss applicable.

\subsection{Second stage: supervised training with pseudo labels}
After training the first-stage model, we use it to classify all chunks generated by the diarization algorithm. For this purpose, an audio segment is only assigned to the target speaker if it is classified as such by the network. This selection process takes place after training, so neither the aggregation function nor the AAM margin is employed during this phase. By following this approach, the second stage can leverage a greater number of relevant audio segments for supervised training, including segments that were previously excluded from VoxCeleb due to a lack of overlap between audio and face data.

\subsection{Exploiting pre-trained models with MHFA}
Training a speaker verification system using our weakly-supervised approach has higher computational requirements compared to fully-supervised methods. This is attributed not only to the two-stage training process, but also to the substantially larger amount of data employed, especially in the first phase, when embedding extractors are trained. 

One effective way to mitigate these costs and enhance performance, without relying on labeled data, is to leverage self-supervised models pre-trained on unlabeled speech. These Transformer-based models are typically trained using a masked speech prediction task as the pre-training objective, which often leads final layers to focus on phonetic information extraction~\cite{hubert}. However, speaker-identity-related tasks may benefit more from low- and mid-level features, which carry most of the distinctive characteristics of a speaker's identity. 

The Multi-Head Factorized Attention (MHFA) \cite{mhfa} addresses this by using two vectors of weights, $\mb{w}^k$ and $\mb{w}^v \in \mathbb{R}^{L}$, to aggregate outputs $\mb{Z}_l \in \mathbb{R}^{T \times F}, l=1,\dots,L$ from all $L$ Transformer layers, where $T$ is the number of frames and $F$ is the feature dimensionality. 
The aggregation and linear projection yield key and value matrices as:
\begin{equation}
    \mb{K} = \left(
        \sum_{l=1}^{L} w_l^k \mb{Z}_l
    \right) \mb{S}^{k}, ~~~
    \mb{V} = \left(
        \sum_{l=1}^{L} w_l^v \mb{Z}_l
    \right) \mb{S}^{v},
\end{equation}
where $\mb{S}^k \in \mathbb{R}^{F \times D}$ and $\mb{S}^v \in \mathbb{R}^{F \times D}$ project features to a $D$-dimensional space.
Employing the learnt query matrix 
$\mb{Q}=[\mb{q}_{1}, \ldots, \mb{q}_{H}] \in \mathbb{R}^{D\times H}$, an attention mechanism variant with $H$ ``heads'' is applied as follows:
\begin{equation}
\mb{c}_h = \mb{V}^T \mr{softmax}(\mb{K} \mb{q}_h),
\end{equation}
where $\mb{c}_{h} \in \mathbb{R}^{D}$ represents the output of each head $h$. The final utterance-level speaker representation is obtained by concatenating these outputs across all heads $\mb{c} = [\mb{c}_{1}^{T},\ldots, \mb{c}_{H}^{T}]^T \in \mathbb{R}^{HD}$. Finally, a linear layer is used to transform $\mb{c}$ to the speaker embedding. 


During our experiments with pre-trained SSL models and the MHFA backend\cite{mhfa}, we saw that we need to do substantially fewer training epochs than training ResNet-based systems from scratch while achieving state-of-the-art results.


\subsection{Utilization of unknown speakers' segments}
Aiming to utilize segments containing speech of speakers different from the target ones (i.e., those not classified as belonging to the target speaker by the model obtained during the first stage), we devise a new training method. It enables training with an extra unknown class and, as a result, allows us to increase the size of the training corpus.

Consider that $N$ is a set of numerical labels of known speakers and $\mc{F}_{\mr{id}}$ maps the example to its corresponding label. For each mini-batch, a second-stage model provides logits $\mb{L} \in \mathbb{R}^{|B| \times |N|}$, with B representing a set of mini-batch examples. Our proposed method extends the logits by adding one class without introducing a new prototype, leading to logits $\mb{L}' \in \mathbb{R}^{|B| \times (|N|+1)}$. For each example $b \in B$ in a mini-batch, we set the logit of the additional class as follows:
\begin{equation}
    \mb{L}'_{b, |N|+1} = 
    \begin{cases}
        0 & \mr{if} ~~ \mc{F}_{\mr{id}}(b) \in N \\
        \frac{1}{|B_N|} 
        \sum_{i \in B_N} \mb{L}_{i, \mc{F}_{\mr{id}}(i)}
        & \mr{otherwise} \\
    \end{cases},
\end{equation}
where $B_N = \{i | i \in B \land \mc{F}_{\mr{id}}(i) \in N \}$ represents a set of examples in a mini-batch with known speaker identities. For examples with known labels, we set the expected logit of the unknown class to zero. Recalling that logits are scaled cosine similarities, zero corresponds to the perpendicularity of extracted embedding and imagined embedding of the other class. Empirically, embeddings that are perpendicular to each other belong to different speakers. In the case of an example for which the label is unknown, we set its logit to an expected logit the network would yield for the target class at a given point of training. We estimate the expectation over one mini-batch. In turn, a loss for the example with an unknown label can be decreased by decreasing the probability of all known classes. Hence, the network is encouraged to model the differences between those known speakers having similar characteristics with the unknown speakers, further increasing the discriminability of the embeddings.

\begin{table}[t]
\caption{Comparison of VoxCeleb2 Development Set Variants}
\begin{center}
\begin{tabular}{lccr}
\toprule
\textbf{Dev. Set} & \textbf{Celebrities} & \textbf{Recordings}  & \textbf{Hours}\\
\midrule
\textbf{Original} & 5,994 & 145,569 & 2,369.0 \\
\textbf{Restricted} & 5,987 & 110,940 & 1,884.3\\
\textbf{Uncut} & 5,987 & 110,940 & 10,211.6  \\
\textbf{Self-labeled} & 5,987 & 110,453 & 6,026.9 \\
\bottomrule
\end{tabular}
\label{tab:voxceleb_data}
\end{center}
\end{table}

\begin{table*}[h!]
\caption{Speaker verification results for the first stage employing ResNet34 model. Results are reported on the three trials of VoxCeleb1. Model denoted as p employs Pyannote diarization.}
\begin{center}
\begin{tabular}{clccccccc}
\hline
\multirow{2}{*}{\textbf{Model}} & \multirow{2}{*}{\textbf{Aggregation /}\boldmath$\tau$} & \multirow{2}{*}{\textbf{Margin}} & \multicolumn{2}{c}{\textbf{VoxCeleb1-O}} & \multicolumn{2}{c}{\textbf{VoxCeleb1-E}} & \multicolumn{2}{c}{\textbf{VoxCeleb1-H}}\\
& & & \textbf{EER} & \textbf{MinDCF} & \textbf{EER} & \textbf{MinDCF} & \textbf{EER} & \textbf{MinDCF} \\  \midrule
m1 & max / - &  \multirow{3}{*}{0.1}  & 4.44 & 0.289 & 4.71 & 0.329 & 7.43 & 0.444\\
m2 & LSE / 0.5 & & 4.33 & 0.262 & 4.33 & 0.278 & 7.01 & 0.403\\
m3 & LSE / 0.5 $\rightarrow$ 0.1 & & 3.85 & 0.243 & 3.90 & 0.250 & 6.30 & 0.364 \\ \midrule
m4 & max / -  &  \multirow{3}{*}{0.0}  & \textbf{3.00} & \textbf{0.187} & \textbf{2.86} & \textbf{0.181} & \textbf{4.68} & \textbf{0.266}\\
m5 & LSE / 0.5 &  & 4.34 & 0.274 & 4.44 & 0.288 & 7.29 & 0.416 \\
m6 & LSE / 0.5 $\rightarrow$ 0.1 &  & 4.86 & 0.315 & 5.20 & 0.333 & 8.25 & 0.462 \\
\midrule
p & max / - & 0.0 &  2.55 & 0.159 & 2.37 & 0.151 & 3.93 & 0.230\\
\bottomrule
\end{tabular}
\label{tab1}
\end{center}
\label{tab:stage_1}
\end{table*}
\section{Experimental Setup}
\subsection{Non-segmented VoxCeleb2 data}
Our training corpus \textit{VoxCeleb2 Uncut dev} has been acquired by downloading full-length audio recordings from YouTube, without their visual components. This approach aimed to retain the complete audio content without segmenting it as it is done for the original VoxCeleb2 dev set, obtaining a larger number of audio hours as shown in Table~\ref{tab:voxceleb_data}. However, as some video clips are no longer available, some speakers have been lost compared to the original set. To enable fair comparisons, we also created a \textit{restricted} version of the dataset, in which the number of speakers and recordings has been limited to match the \textit{uncut} version.

As it is also reflected in Table~\ref{tab:voxceleb_data}, the pseudo labels generated by the best first-stage trained ResNet model (\textit{m4} from Table\ref{fig:stage1}) used for training the fully-supervised second stage cover approximately three times more audio hours than the original dataset. This substantial increase in training data is expected to enhance the robustness and overall performance of the fully-supervised model.




\subsection{Training configuration}
\label{sec:training_config}
For better reproducibility and straightforward experimentation, we integrated our approach of weakly-supervised training into the WeSpeaker toolkit~\cite{wang2024advancing}. The main contributions to the toolkit are the data loader and the versions of AAM with aggregation. Notably, due to the nature of the data partitioned into multiple clusters, our data-loading implementation provides mini-batches of dynamic size fluctuating up to $\pm$10\% around the target mini-batch size, which should be kept in mind w.r.t. GPU resources.

As the backbone of our embedding extractor, we use ResNet34 and MHFA along with WavLM features. For ResNet34, we employ 2D convolutional layers with 64, 128, 256, and 256 filters. The original block structure has been modified following \cite{identity_mappings}. 
Since our diarized segments may contain non-speech content, we employ instant normalization instead of batch normalization \cite{instance_norm}. The input to our ResNet models consists of 400 frames of 80-dimensional log Mel-filter bank energy (fbank) features.

In experiments employing pre-trained models, we opt for a WavLM Base+ \footnote{\url{https://huggingface.co/microsoft/wavlm-base-plus}} for its size and performance of downstream models that build on top of it in the SUPERB benchmark~\cite{yang21_superb}. The input duration is 4\,s. The MHFA back-end comprises 64 query heads and provides 256-dimensional embeddings.

For AAM, we employ a scale $s=30$ for all models. Given the uncertainty of segments belonging to the target celebrity, we experimented with incorporating margin during the first stage, as it will be detailed in the experiments section. For the second stage, we scheduled the margin from 0.1 to 0.3 or 0.2 for ResNet34- and MHFA-based models, respectively, following our previous empirical experience. The models are trained employing SGD with a momentum of 0.9. Following an initial warm-up phase, the learning rate is decayed exponentially. The maximum learning rates are set to 0.2 for ResNet34-based models and 0.01 for MHFA-based models, with final learning rates of $5\times 10^{-5}$ and $4.4\times 10^{-3}$, respectively.

\section{Experiments}

\subsection{First stage}
During the first stage, we experimented training with three aggregation functions: max-pooling (model \textit{m1}), log-sum exp (LSE) with fixed $\tau$ (\textit{m2}) and LSE with scheduled $\tau$ (\textit{m3}), employing a margin of 0.1. As presented in Table~\ref{tab:stage_1}, best results are obtained employing the LSE aggregation function, particularly when the $\tau$ parameter is scheduled, confirming earlier findings \cite{stafylakis22_interspeech}.

While previous work primarily aimed to establish a viable pipeline for training with weak labels, it lacked an in-depth hyperparameter exploration, such as the use of margin and its effects. Enforcing a margin during training with strong labels (such as VoxCeleb) helps to make speaker clusters in the embedding space more separable. However, in our case, the examples do not contain only the target speaker's speech, which makes the task more difficult. Then, enforcing the margin, which makes training more challenging, is questionable. It turns out that releasing the margin constraint increases performance when using the max-pooling (\textit{m4} in Table~\ref{tab:stage_1}), achieving our best results. We observe an interesting trade-off between the two hyperparameters: margin enforcement benefits from the smoothing effect of LSE (\textit{m2} and \textit{m3}), while removing the margin favors stricter pooling strategies such as max-pooling (\textit{m4}).

\begin{table}[t!]
\caption{Effects of employing WavLM features along with MHFA for training the embedding extractor during the first stage.}
\begin{center}
\begin{tabular}{clcccc}
\hline
\textbf{Model} & \textbf{Aggregation /}\boldmath$\tau$ & \boldmath$m$  & \textbf{Vox1-O} & \textbf{Vox1-E} & \textbf{Vox1-H}\\
\hline
m7 & max / - &  \multirow{2}{*}{0.1} & \textbf{1.25} & \textbf{1.36} & \textbf{2.55} \\
m8 & LSE / 0.5 &  & 3.82 & 3.71 & 5.86\\ 
\hline
m9 & max / - &  \multirow{2}{*}{0.0} & 1.51 &  1.76 & 3.36 \\ 
m10 & LSE / 0.5 &  & 3.59 & 3.10 & 5.60\\ 
\hline
\end{tabular}
\label{tab:MHFA_first_stage}
\end{center}
\end{table}


Additionally, Table~\ref{tab:MHFA_first_stage} presents the impact of using features from the pre-trained WavLM model and the MHFA back-end. This approach outperformed the use of ResNet34 as the backbone architecture, keeping a similar training time, showing that the proposed methodology is model-agnostic and can be effectively integrated with self-supervised models. Models employing max-pooling aggregation lead to superior performance than those using LSE, with the improvement being even more pronounced in this case. In contrast to the results observed with ResNet34, max-pooling aggregation benefits from adding margin (\textit{m7} vs \textit{m9}) to the AAM loss. We hypothesize that the WavLM model provides speaker-enriched features resulting from pre-training that allow for stricter loss. Although we achieve more competitive results with WavLM + MHFA, they are still inferior compared to strong supervision training, as shown in Table \ref{tab:second_stage}.

\subsection{Evaluation of method validity}
In this section, we evaluate the efficacy of our proposed method for training an embedding extractor employing weakly-labeled data. Previous experiments employed a basic BIC-based HAC algorithm to generate input segments, which tend to overestimate the number of clusters within each recording. To assess the dependence of method performance on segment quality, we train with segments obtained with a state-of-the-art pre-trained diarization model Pyannote.audio 3.1 \cite{Bredin23, Plaquet23}. This advanced diarization system not only provided segments with higher speaker purity, removing segments containing only noise and music, but also allowed us to constrain the maximum number of speakers to 4. 
Although the higher-quality segments led to improved results (model \textit{p} in Table~\ref{tab:stage_1}) the gains were not as significant as expected. This indicates that our method is robust and capable of extracting meaningful speaker embeddings even from lower-quality segments generated by the baseline diarization approach. Importantly, unlike the Pyannote system, our method does not require any training for segmentation, offering a simpler and more accessible alternative in practice.

To further assess the validity of our method, we compare the quality of segments selected by our best first-stage model (\textit{m7}) w.r.t. the semi-oracle ones. Since ground-truth annotations are not available for segments provided by the baseline diarization, we employed state-of-the-art ResNet293 trained in a supervised way on the VoxCeleb2 dataset (using the Wespeaker toolkit) to provide semi-oracle labels. As demonstrated in Table~\ref{tab:seg_select}, both the semi-oracle model and our best trained system selected utterances from all 5987 available speakers. In terms of number of hours, there is only a loss of 42 hours from the proposed model compared to the semi-oracle system. We also confirm a high overlap of segments as precision and recall w.r.t. the semi-oracle ones are 94.16 and 93.68, respectively. These findings support the validity of our method.
\begin{table}[t!]
\caption{Segments selection employing embedding extractor from semi-oracle model and our best weakly supervised first-stage model}
\begin{center}
\begin{tabular}{ccccc}
\toprule
\textbf{Segments} & \textbf{Speakers} & \textbf{Duration (h)} & \textbf{Precision} & \textbf{Recall}\\
\midrule
\textbf{Semi-oracle} & 5987 & 6304.5 & reference & reference\\
\textbf{Proposed (m7)} & 5987 & 6262.1 & 94.16 & 93.68\\ 
\bottomrule
\end{tabular}
\label{tab:seg_select}
\end{center}
\end{table}

\begin{table*}[t!]
\caption{Results for the second stage. Margin was scheduled from 0.1 to 0.3 for every experiment. When pseudo-labels are used for training, data column refers to the first-stage model used for segment selection. For minDCF, target trial probability is set to 0.05, assuming equal costs for misses and false alarms. References \cite{miara24_interspeech} and \cite{bingTASL} report results using a prior of 0.01.}
\begin{center}
\begin{tabular}{ccclcccccc}
\toprule
\multirow{2}{*}{\textbf{Supervision}} & \multirow{2}{*}{\textbf{Data}} &  \multirow{2}{*}{\textbf{Speakers}} &  \multirow{2}{*}{\textbf{Model}} & \multicolumn{2}{c}{\textbf{Vox1-O}} & \multicolumn{2}{c}{\textbf{Vox1-E}} & \multicolumn{2}{c}{\textbf{Vox1-H}} \\
&&&& \textbf{EER} & \textbf{minDCF} & \textbf{EER} & \textbf{minDCF} & \textbf{EER}& \textbf{minDCF} \\
\midrule
\multirow{3}{*}{strong} & \multirow{3}{*}{original} & \multirow{3}{*}{5,994} & ResNet34 & 0.87 & -- & 1.05 & -- & 1.96 & --\\ 
 &  & & ResNet293  & 0.56 & -- & 0.76 & -- & 1.43 & -- \\
 & & & WavLM + MHFA & 0.86 & 0.067 & 0.88 & 0.056 & 1.77 & 0.107 \\
\midrule
\multirow{3}{*}{strong} & \multirow{3}{*}{restricted} & \multirow{3}{*}{5,987} & ResNet34  & 1.11 & 0.064 & 1.13 &  0.073 & 2.07 & 0.121 \\ 
 &  & & ResNet152 & 0.69 & 0.048 & 0.82 & 0.052 & 1.57 & 0.092 \\
 &  & & WavLM + MHFA & 0.86 & 0.065 & 0.86 & 0.057 & 1.78 & 0.111\\ 
\midrule
\multirow{8}{*}{pseudo-labels} 
 & \textit{m3} & 5,945 & \multirow{2}{*}{ResNet34} & 0.97 & 0.056 & 1.06 & 0.066 & 1.88 & 0.111 \\ 
&  \textit{m4} & 5,987 & & 0.99 & 0.054 & 1.02 & 0.065 & 1.83 & 0.107 \\ 
&  \textit{m4} & 5,987 & ResNet152 & 0.84 & \textbf{0.051} &\textbf{0.85} & \textbf{0.053} & \textbf{1.55} & \textbf{0.091} \\ 
& \multirow{2}{*}{\textit{m7}} &  \multirow{2}{*}{5,987} & WavLM + MHFA & \textbf{0.83} & 0.062 & 0.92 & 0.062 & 1.91 & 0.121 \\
&&& ~~+ unknown class & 0.83 & 0.061 & 0.92 & 0.058 & 1.86 & 0.117 \\
\cmidrule(lr){2-10}
& \cite{stafylakis22_interspeech} & 5,935 & ResNet34 &  1.58 & 0.103 & 1.82 & 0.118 & 3.23 & 0.184 \\ 
& \cite{bingTASL} & 5,994 & ECAPA-TDNN & 1.57 & 0.167 & 1.88 & 0.196 & 3.29 & 0.294\\
& \cite{miara24_interspeech} & 5,994 & WavLM + MHFA & 0.99 & 0.091 & 1.21 & 0.126 & 2.35 & 0.221 \\

\bottomrule
\end{tabular}
\label{tab:second_stage}
\end{center}
\end{table*}

\subsection{Second Stage}
In Table~\ref{tab:second_stage}, we present a comparative analysis of our second-stage systems against state-of-the-art models. These second-stage systems were trained using self-labeled data derived from our best-performing first-stage models (\textit{m3}, \textit{m4}, \textit{m7}).
We compared our results with strong supervised baselines, including ResNet34 and ResNet293 models trained on the original VoxCeleb2 dataset, following the WeSpeaker recipe with their proposed tuned hyperparamenters\footnote{WeSpeaker Official VoxCeleb results: \url{https://github.com/wenet-e2e/wespeaker/tree/master/examples/voxceleb/v2}}. 

To ensure a fair comparison in terms of the number of speakers, we also trained the same architectures as our proposed models (ResNet34, ResNet152, and WavLM + MHFA) on the \textit{restricted} subset of VoxCeleb2, using the same hyperparameter settings described in Section~\ref{sec:training_config}. Our models trained with pseudo-labels perform much better than their counterparts trained during the first stage, and also outperform the best second-stage model reported in our previous work \cite{stafylakis22_interspeech}. When it comes to ResNet34 backbone, training with pseudo-labels generated by models \textit{m3} and \textit{m4} leads to similar results, despite the loss of some speakers when selecting segments with \textit{m3}. Both of them outperform fully-supervised training on the \textit{restricted} set, and achieve results comparable to the official WeSpeaker results, even outperforming the VoxCeleb1-H subset. Therefore, 
our approach demonstrates that utilizing weakly labeled data increases the quantity of high-quality pseudo labels, which in turn enhances the effectiveness of training in the second stage.

We show that strong results are also obtained with the model based on WavLM and MHFA. The benefit of such a system is that it is trained approximately four times faster compared to the ResNet34-based model.
Due to the limited capability of our servers, we could not train ResNet293 from scratch. Despite not outperforming its SOTA results employing the full VoxCeleb2, we obtain very good results employing the smaller ResNet152. This is particularly noteworthy given the loss of data, including seven speakers and additional recordings from other individuals, as detailed in Table~\ref{tab:seg_select}.

Finally, Table~\ref{tab:second_stage} also includes preliminary results using an approach that incorporates unlabeled segments during second-stage training. Here, we introduced an "unknown" class in the middle of the training to model unseen speakers. We followed a specific selection process with filtering to focus on difficult segments of unknown speakers. If the target speaker was among the 10-best predictions of the first-stage model, we discarded a corresponding segment, as it may contain a target speaker's speech (and it would be misleading when considered as belonging to the unknown class). Out of the remaining segments that do not correspond to the target speaker, we selected 5\% with the highest LSE of logits not to overweigh examples of a newly added class. This method further improved WavLM + MHFA performance, particularly on the VoxCeleb1-H evaluation set.

\subsection{Comparison with self-supervised models}
We compared our approach with recent state-of-the-art self-supervised methods that rely on segmented VoxCeleb unlabeled data. These methods typically follow a two-stage pipeline, where a speaker encoder is first trained using DINO-based objectives and then fine-tuned using pseudo-labels derived from the initial representations. Specifically, we evaluated against a WavLM-MHFA approach trained with pseudo-labels obtained by clustering DINO speaker embeddings \cite{miara24_interspeech}, as well as the dynamic loss-gate and label correction (DLG-LC) method \cite{bingTASL}, which incorporates dynamic label correction mechanisms.

Despite only relying on source VoxCeleb audio without segmentation via video-based intersection, as shown in Table~\ref{tab:second_stage} our method not only outperforms these approaches \cite{miara24_interspeech,bingTASL} but also improves fully-supervised training, an achievement not yet reported by existing self-supervised speaker recognition models. The comparison between both WavLM+MHFA approaches (\textit{m7} vs \cite{bingTASL}) highlights the advantages of using raw, unsegmented audio, as removing video-based constraints enables leveraging significantly more data, which directly translates into improved performance.




\section{Conclusions}
In this paper, we advanced our previous work on training speaker verification embedding extractor using only recording-level speaker labels. Our pipeline re-implementation employing WeSpeaker toolkit achieved significant performance gains compared to previously reported ResNet results,  outperforming fully-supervised models trained with the same number of speakers. We further enhanced our results by exploiting SSL models, employing WavLM along with a  MHFA backend, which also allowed us to decrease training time.

We demonstrated that our embedding extractor's segment selection performance closely matched a semi-oracle model. Additionally, varying the quality of segments employed for training the embedding extractor led to minimal degradation, showing the robustness of our method. These findings confirm the validity of our approach, evidencing the benefits of exploiting the whole audio recording from the source of VoxCeleb.

Most notably, our approach unlocks the potential to train directly on complete, unsegmented audio recordings annotated only with recording-level speaker labels. This eliminates the need for time-consuming segment-level labeling and makes it practical to leverage large-scale datasets with minimal supervision.


\bibliographystyle{IEEEtran}
\bibliography{bibliography}
\end{document}